\documentclass[a4paper,11pt]{article}
\title{\vspace{-2.5cm} \bf Gravitational collapse of thin shells of dust\\in asymptotically flat Shape Dynamics}
\author{Flavio Mercati$^1$\footnote{\href{mailto:flavio.mercati@gmail.com}{flavio.mercati@gmail.com}}, Henrique Gomes$^1$, Tim Koslowski$^{2,3}$ and Andrea Napoletano$^4$\vspace{12pt}\\
\it \small $^1$Perimeter Institute for Theoretical Physics, 31 Caroline Street North,
\\
\it \small Waterloo, ON, N2L 2Y5 Canada,\\
\it \small $^2$University of New Brunswick, Fredericton, NB, E3B 5A3 Canada  \\ \it \small $^3$Instituto de Ciencias Nucleares, Universidad Nacional Aut\'onoma de Mexico,
\\
\it \small D.F. 04510, M\'exico,\\
\it \small $^4$Dipartimento di Fisica, Universit\`a di Roma La Sapienza,\\
\it \small  P.le Aldo Moro 2, 00185 Roma, Italia}
\date{\today}

\usepackage[left=1in,right=1in,top=1in,bottom=1in]{geometry} 
\usepackage{amssymb,bbold}
\usepackage{amsmath}
\usepackage[usenames,dvipsnames]{color}
\usepackage{tensor}
\usepackage{bbm}
\usepackage{graphicx}
\usepackage{cancel}
\usepackage{mathrsfs}

\usepackage{hyperref}
\hypersetup{
    colorlinks=true,         
    linkcolor=red,          
    citecolor=red,        
    urlcolor=Violet             
}

\usepackage[normalem]{ulem}
\usepackage{mathrsfs}
\usepackage[mathscr]{euscript}

\def\Xint#1{\mathchoice
   {\XXint\displaystyle\textstyle{#1}}%
   {\XXint\textstyle\scriptstyle{#1}}%
   {\XXint\scriptstyle\scriptscriptstyle{#1}}%
   {\XXint\scriptscriptstyle\scriptscriptstyle{#1}}%
   \!\int}
\def\XXint#1#2#3{{\setbox0=\hbox{$#1{#2#3}{\int}$}
     \vcenter{\hbox{$#2#3$}}\kern-.5\wd0}}

\def\dashint{\Xint-}

\newcommand{\sfrac}[2]{\ensuremath{\textstyle{\frac{#1}{#2}}}}
\newcommand{\st}[1]{\text{\tiny \rm #1}}

\newcommand{\mass}{m}

\renewcommand{\d}{{\rm d}}

\begin{document}

\maketitle

\begin{abstract}
In a recent paper, one of us studied spherically symmetric, asymptotically flat solutions of Shape Dynamics, finding that the spatial metric has characteristics of a wormhole --  two asymptotically flat ends and a minimal-area sphere, or `throat', in between. In this paper we investigate whether that solution can emerge as a result of gravitational collapse of matter. With this goal,  we study the simplest kind of spherically-symmetric matter: an infinitely-thin shell of dust. Our system can be understood as a model of a star accreting a thin layer of matter.
We solve the dynamics of the shell  exactly and find that, indeed, as it collapses, the  shell leaves in its wake the wormhole metric. 
In the maximal-slicing time we use for asymptotically flat solutions, the shell only approaches the throat asymptotically and does not cross it in a finite amount of time (as measured by a clock `at infinity'). This leaves open the possibility that a more realistic cosmological solution of Shape Dynamics might see this crossing happening in a finite amount of time (as measured by the change of relational/shape degrees of freedom).
\end{abstract}

\section{Introduction}

Shape Dynamics (SD) is a Hamiltonian theory which describes gravity as the evolution of a 3D conformal geometry.  In this description, 4D spacetime is not the fundamental dynamical object of the theory, and must be understood as an emergent concept. Nonetheless it can be useful to describe the behaviour of weakly-backreacting `probes'. The utility of the space-time picture comes about from the relation of the two theories: 
SD is equivalent to the Hamiltonian formulation of General Relativity (GR)~\cite{ADM} for solutions of the latter which possess a complete \lq{}constant mean curvature\rq{} foliation.
%
%
%
This slicing condition fails at the global level, already for simple examples: Sch\-warz\-schild's spacetime does not possess a complete CMC slicing. In~\cite{Birkhoff_SD}, one of us found that, respecting  spherical symmetry and standard asymptotic flatness conditions, a solution of the equations of SD could be built which covers only the two \emph{non-singular} quadrants of the Kruskal extension of  Sch\-warz\-schild.  The spatial metric is that of a wormhole, possessing a `throat '(a minimal-area sphere) and two asymptotic ends. At the throat the spatial metric is smooth, but the spacetime metric  has a defect. This  discrepancy  is possible in spite of Birkhoff\rq{}s theorem, because in SD it is the spatial conformal geometries that need to be regular, whereas in GR it is the 4D spacetime geometry that needs to be well-bahaved. 
%
%

The result of~\cite{Birkhoff_SD} was nonetheless preliminary:  the assumption of spherical symmetry leaves no room for degrees of freedom of the conformal geometry (the Cotton tensor vanishes~\cite{BirkhoffFlavio}). Moreover~\cite{Birkhoff_SD}  did not address the issue of whether this solution would form from gravitational collapse, \emph{i.e.} whether it was physical or not. 
Here we partially remedy both shortcomings in a minimal way,  by considering spherically symmetric matter sources, which introduce genuine dynamical degrees of freedom for the shell,\footnote{Albeit not genuine ``shape'' degrees of freedom: for those one would need to introduce more than one shell, to form scale-invariant ratios of their variables.}  and studying their gravitational collapse. 

As was done in most recent works in SD~\cite{Birkhoff_SD,BirkhoffFlavio,ThroughTheBigBang},  we exploit the local equivalence between SD and GR in CMC foliation to simplify the calculations. The input from SD is limited to: i) insisting that the spatial conformal geometries remain regular throughout evolution, and 
 ii) neglecting any regularity requirement on the spacetime metric.
 


There are other flaws in  \cite{Birkhoff_SD} which we are also not going to address in this paper. The first is the assumption of asymptotic flatness; according to the relational underpinnings of SD, the theory should be  based on a closed spatial manifold. The second is the boundary conditions at infinity, which have been  borrowed wholesale from GR, while in SD they should be set by the physical behaviour of matter `at infinity'~\cite{BirkhoffFlavio}. The last one is the  use of maximal-slicing time, which violates one of the relational pillars on which Shape Dynamics rests: that time should be derived from the change of physical (\emph{i.e.} shape, or conformally-invariant) degrees of freedom~\cite{FlaviosSDtutorial}.  Nonetheless, there are arguments that show that, as far as the solution exists, it will reliably represent a``background experienced'' space-time for weakly back-reacting matter degrees of freedom \cite{Tim_Proceedings_TheoryCanada9}, and maximal-slicing time should approximate the amount of change experienced by a clock far away from the origin. In sum, the goal of the present paper is to study whether the solution found in~\cite{Birkhoff_SD} can emerge as the result of gravitational collapse, and therefore we postpone addressing these issues to further studies.

Differently from previous work, here we need to couple matter -- pressureless dust in particular -- in a first attempt to model gravitational collapse. The simplest distribution of dust which respects spherical symmetry is an infinitely thin sphere.
The coupling of SD to matter is borrowed (for phenomenological reasons) from GR, by working, again, in the gauge in which the two theories are equivalent. This is done in Section \ref{sec:dust}. Before that, we will solve the SD equations in vacuum in Section \ref{sec:vacuum}, and then insert those constructions where appropriate when treating the full  system coupled to dust. The full system is described in Sec.~\ref{sec:fullsystem}, where the reduced phase space of physical degrees of freedom is characterized, and the on-shell orbits describing the evolution of the collapsing shell are found. Sec.~\ref{sec:conclusions} contains an outlook of the result.

\section{Vacuum spherically symmetric solutions}\label{sec:vacuum}

The constraints of  Shape Dynamics, in the gauge in which it is equivalent to 
maximal-slicing GR, are~\cite{gryb:shape_dyn,Gomes:linking_paper,FlaviosSDtutorial}
\begin{equation}\label{ADMconstraints} \textstyle
\mathcal H  = 
 \frac{1}{\sqrt g} \left( p^{ij} p_{ij} - {\frac 1 2} p^2 \right) - \sqrt g \, R 
 \,, ~~
\mathcal H_i  = -2 \, 
\nabla_j p^j{}_i  
\,, ~~
\mathcal C =  p \,,
\end{equation}
where $g_{ij}$ is the spatial metric,  $p^{ij}$ its conjugate momentum and $p = g_{ij} p^{ij}$. These constraints need to be valid at all times. The time evolution of the fields is generated by the following equations:
\begin{equation}\label{EqsOfMotion}
\begin{aligned}
\dot g_{ij} =& \textstyle \frac{2 N}{\sqrt g} \left( p_{ij}- {\sfrac 1 2} g_{ij} p \right) + \nabla_i \xi_j + \nabla_j \xi_i \,,
\\
\dot p^{ij} =&  \textstyle - N \sqrt g \left( R^{ij} - {\sfrac 1 2} g^{ij} R  \right) 
+ \frac{N}{2\sqrt g} g^{ij} \left( p^{k\ell} p_{k\ell} - {\sfrac 1 2} p^2\right) 
\\&\textstyle - \frac{2 N}{\sqrt g} \left( p^{ik} p_k{}^j - {\sfrac 1 2} p \, p^{ij} \right) 
+ \sqrt g \left( \nabla^i \nabla^j N - g^{ij} \Delta N \right) 
\\ &\textstyle + \nabla_k (p^{ij} \xi^k) - p^{ik} \nabla_k \xi^j - p^{kj} \nabla_k \xi^i  \,,
\end{aligned}
\end{equation}
which depend on a Lapse function $N$ and a shift vector $\xi_i$. The latter is fixed by our choice of coordinates, while the former is fixed by the requirement of preservation in time of the conformal constraint $\mathcal C \approx 0$. This gives rise to the so-called lapse fixing equation:
\begin{equation}\label{LFE}
{\sfrac 1 2} \sqrt g N R - 2 \sqrt g  \Delta N + {\sfrac 3 2} \frac {p^{ij} p_{ij}}{\sqrt g} N = 0 \,.
\end{equation}

Using the notation of~\cite{BirkhoffFlavio}, the most generic spherically symmetric metric and momentum can be written as:
\begin{equation} \label{SphericalSymmetryAnsatz}
\textstyle
g_{ij} = \text{diag} \, \left\{  \mu^2 , \sigma  , \sigma \, \sin^2 \theta  \right\}, 
p^{ij} =  \text{diag} \, \left\{  \frac{f}{\mu} , {\sfrac s 2}     ,  \frac{s}{2 \sin^2 \theta}  \right\} \sin \theta \,,
\end{equation}
where $\mu$, $\sigma$, $f$ and $s$ are functions of the radial coordinate $r$ only.
There is an analogue ansatz for the shift vector: $\xi^i = (\xi (r) , 0 ,0 )$.

\subsection{Solution of the constraints}

Replacing the ansatz~(\ref{SphericalSymmetryAnsatz}) into  the constraints (\ref{ADMconstraints}), we get
\begin{equation}\label{SphericalADMconsts}
\begin{aligned}
& \sigma ^2 \mu s^2 +  4 f^2 \mu ^3 -4 f \sigma \mu ^2 s +   12 \sigma \mu \sigma '' \\
&  - 12 \sigma \sigma ' \mu ' - 3 \mu (\sigma')^2 - 12 \sigma \mu ^3  = 0 
\\
& \mu f' - {\sfrac 1 2} s \sigma' = 0 \,, \qquad \qquad  \mu f + s \sigma = 0 \,, 
\end{aligned}
\end{equation}
where $'$ refers to the radial derivative $\frac \partial {\partial r}$. The last equation can be solved algebraically, $s  = 
 - \frac{\mu}{\sigma} \, f$,  and after replacing  this expression for $s$ in~(\ref{SphericalADMconsts}), it is easy to see that the diffeomorphism constraint can be written as a total derivative, $\frac{\mu}{\sqrt \sigma} \left(  f \sqrt \sigma 
 \right)'  = 0$. The solution of this equation is
\begin{equation}\label{SolutionVacuumDiffConstraint}
 f = 
  \frac{A}{\sqrt \sigma}\,,
\end{equation}
where $A$ is an integration constant (meaning that it is spatially constant but can, in principle, still be a function of time). Finally, with a little work one can check that the Hamiltonian constraint can be rewritten as
\begin{equation}
\textstyle
- \frac{\sigma^{\frac 1 2} \mu}{\sigma'} \frac{\partial}{\partial r} \left[ \left(\frac{\sigma'}{\sigma^{\frac 1 4} \mu}\right)^2 - 4 \sqrt{\sigma} - \frac{f^2}{\sqrt \sigma} 
 \right] = 
  \frac{2 \, f \,\mu}{\sigma'} \left( f' + {\sfrac 1 2} \, f \, \frac{\sigma'}{\sigma} 
\right) \,,
\end{equation}
where the term $\left( f' + {\sfrac 1 2} \, f \, \frac{\sigma'}{\sigma} 
\right)$ is identical to the diffeomorphism constraint and therefore vanishes on-shell. The remaining term is a total derivative, and we can solve the equation by introducing a new integration constant $m$  (the $-4$ factor is introduced for convenience),
\begin{equation}\label{DefinitionIntegrationConstant_m}
\frac{(\sigma')^2}{\sqrt{\sigma} \mu^2} - 4 \sqrt{\sigma} - \frac{f^2}{\sqrt \sigma}  
= - 8 \, \mass \,.
\end{equation}
Replacing the solution~(\ref{SolutionVacuumDiffConstraint}), 
we get a relation between $\sigma$ and $\mu$, and since the latter appears without derivatives, the easiest thing is to solve with respect to $\mu$:
\begin{equation}\label{VacuumSolutionForMu}
\mu^2 =  \frac{(\sigma')^2}{\frac{A^2}{\sigma }  - 8 \, \mass  \,\sqrt{\sigma}
+ 4 \, \sigma 
 } \,.
\end{equation}

We found a solution to all our constraints which apparently holds for any choice of the remaining free function $\sigma(r)$. This is a reflection of radial diffeomorphism invariance. In fact notice how $g_{rr} = \mu^2$ is homogeneous of degree two in $\sigma'$: the expression $\mu^2 \d r^2 \propto (\sigma' \d r)^2$ appearing in the metric is explicitly invariant under changes of radial coordinate.
However the choice of $\sigma(r)$ is not completely arbitrary. If we require regularity of the conformal geometry, there are obstructions to the values that $\sigma$ can take. In fact, by inspecting~(\ref{VacuumSolutionForMu}) we can see how the right-hand side is not guaranteed to be positive. It relies on the following fourth-order polynomial of $\chi = \sqrt{\sigma}/m$
\begin{equation}\label{MordorPolynomial}
\mathscr P (\chi) = C^2 
- 2 \,\text{sign}(\mass)\, \chi^3  + \chi^4 \,, ~~~ C = \frac{A}{2 \, \mass^2} \,.
\end{equation}
See~\cite{BirkhoffFlavio} for a detailed discussion of the roots of $\mathscr P$. Here we only need to observe that if $C=0$ $\mathscr P$ is positive when $\chi > 2$.

\subsection{Solution of the equations of motion}

The equations of motion require previous calculation of the lapse from Eq.~(\ref{LFE}).
Under the assumption of spherical symmetry  (which for a scalar function like the lapse is just $N=N(r)$), the lapse-fixing equation reduces to
\begin{equation}\label{SphSymmLFE1}
\begin{aligned}
&\!\!\!\!\!  \left(\frac{4 f     s }{\mu   \sigma  }  -\frac{4 f ^2   }{\sigma  ^2} -\frac{4    \mu'  \sigma' }{\mu^3 \sigma  }+\frac{4    \sigma'' }{\mu  ^2 \sigma  }-\frac{   (\sigma')^2}{\mu^2 \sigma^2}-\frac{4   }{\sigma}-\frac{   s^2}{\mu^2} \right)  N 
\\
&\!\!\!\!\! 
- \left( \frac{8 \mu'}{\mu  ^3}+\frac{8 \sigma' }{\mu^2 \sigma  }\right) N' + \frac{8 N'' }{\mu^2} = 0\,.
\end{aligned}
\end{equation}
The solution is then $N = c_1 \, N_1 + c_2 \, N_2$, a linear combination of the two linearly independent solutions, which are
\begin{equation}\label{SolutionLFEtwinshell}
N_1 = \frac{\sigma'}{2 \mu \sqrt{\sigma}} \,,
\qquad
N_2 =  \frac{\sigma'}{2 \mu \sqrt{\sigma}} \dashint \frac{\mu^3}{(\sigma')^2} \d r    \,,
\end{equation}
where the symbol $\dashint$ refers to the principal value integral, which is needed because its argument contains the term
\begin{equation}
\frac{\mu^3}{(\sigma')^2}  = \frac{|\sigma'|}{\left(\frac{A^2}{\sigma }   - 8 \, \mass  \, \sqrt{\sigma}  + 4 \, \sigma\right)^{3/2}} \,,
\end{equation}
which diverges when $\sigma$ approaches a zero of $\mathscr P$ (which has to be an extremum of $\sigma$~\cite{BirkhoffFlavio}). This divergence has opposite sign on the two sides of the extremum (the left- and right- limite are opposite), and the degree of divergence is the same, so that the following quantity is finite:
\begin{equation}
\!\!\!
\dashint_{r_1}^{r_2}\frac{\mu^3 \d r}{(\sigma')^2}    = \lim_{\epsilon \to 0} \left( \int^{\tilde r - \epsilon}_{r_1} \frac{\mu^3 \d r}{(\sigma')^2} + \int_{\tilde r + \epsilon}^{r_2} \frac{\mu^3 \d r}{(\sigma')^2} \right) \,,
\end{equation}
where $\tilde r \in ( r_1 , r_2)$ is the point where $\sigma$ has its extremum.

Once we have the lapse we can calculate the equations of motion for the metric, the first of Eqs.(\ref{EqsOfMotion}).
Using the spherical symmetry ansatz we get that the $\dot g_{\theta\theta}$ and  $\dot g_{\phi\phi}$  equations are identical, and completely fix the shift vector:
\begin{equation}\label{SolEquation_gdotrr}
\xi_i = \delta^r{}_i  \left( f \, N + \dot \sigma \right) /\sigma' \,.
\end{equation}
Replacing the above solution of $\xi_i$ in the  $\dot g_{rr}$ equation (as well as the solutions of the ADM constraints), we find that they depend nontrivially on the lapse. Fortunately, replacing the solution of the lapse fixing equation~(\ref{SolutionLFEtwinshell}), we find that the two principal-value integrals never appear explicitly -- they always have an $r$-derivative acting on them, and we can use the fact that
\begin{equation}
\!\!\! \partial_r \dashint \frac{\mu^3\d r}{(\sigma')^2}  = \frac{\mu^3}{(\sigma')^2} \,, ~~ \partial_r  \dashint \frac{\sigma^{3/2} \mu^3\d r}{(\sigma')^2}  = \frac{\sigma^{3/2} \mu^3}{(\sigma')^2} \,,
\end{equation}
to simplify the expression. The equation then reduces to
\begin{equation}
   - 96 \,\dot \mass \, \sigma^{3/2}  + 6 A  \left(2 \dot A + c_2\right)= 0\,.
\end{equation}
In order for the above equation to hold for any choice of $\sigma(r)$ the only possibility is that 
\begin{equation}\label{SolutionEqOfMotionThinShell}
c_2 = -2\,\dot A \,, \qquad \dot \mass  = 0 \,.
\end{equation}
We fixed one of the two integration constant present in the lapse, $c_2$. $c_1$ cannot be fixed because the system is re\-para\-me\-tri\-za\-tion-invariant, which implies the freedom to specify the value of the lapse at a point. Moreover, we discovered a conserved quantity: the integration constant $\mass$.  This quantity has the significance of `mass-energy', and it is conserved because the system is spherically symmetric, so it cannot radiate its energy away in the form of gravitational waves, and a form of conservation of energy similar to what holds in field theory on Minkowsi spacetime holds. The quantity $\mass$ is actually what the \emph{Misner--Sharp mass} reduces to in vacuum~\cite{BirkhoffFlavio,MisnerSharpMass}. 
%

The equations of motion for the momenta, the second of Eq.~(\ref{EqsOfMotion}), are
identically satisfied if one imposes the conditions (\ref{SolutionEqOfMotionThinShell}), and therefore add no further information. We have then been able to solve exactly both the constraint equations and the equations of motion in vacuum. This is a significant result that can be exploited to build dynamically meaningful solutions of Shape Dynamics, for example by using some localized spherically symmetric distribution of matter, which leaves most of space empty. 

\section{Coupling a thin shell of dust}\label{sec:dust}

A thin spherically-symmetric shell of dust is just the continuum limit of a homogeneous distribution of particles on a sphere, each one of which move radially with the same speed.
One can deduce the appropriate contribution of such a source to the constraints of Shape Dynamics from that of a point particle.

\subsection{New form of the constraints}

The Hamiltonian and diffeomorphism constraints of ADM gravity coupled to a massive point particle are
\begin{equation}\!\!
\begin{aligned}
&\textstyle  \mathcal  H =
  \delta^{(3)}(x^i- y^i) \sqrt{g^{ij} \, p_i \, p_j +m_0^2}  \,,
\\
& \mathcal  H_i =
\delta^{(3)}(x^i- y^i) \, p_i  \,,
\end{aligned}
\end{equation}
%
%
%
where $y^i$ are the coordinates of the particle, $p_i$ its momentum and $m_0$ its rest mass. The above constraints can be straightforwardly
derived from the standard Einstein--Hilbert action coupled to a point particle.
Note that $p_i$ is included as a cotangent vector, and this arises from minimal coupling.  It is not hard to show that the constraints above are first-class.

Now take a uniform distribution of point particles on the surface of a sphere of radius $R_\st{S}$, and take the continuum limit. The  constraints become
\begin{equation}
\begin{aligned}
 & \mathcal  H =
  \sqrt h  \, \rho(R_\st{S}) \, \delta(r-R_\st{S}) \sqrt{g^{rr} \, p_r^2 +m_0^2 } \,, 
\\
& \mathcal  H_i =
 \delta^r{}_i \, \sqrt h \, \rho(R_\st{S}) \, \delta (r-R_\st{S}) \, p_r \,,
\end{aligned}
\end{equation}
where $h_{ab}$ is the metric induced on the sphere by $g_{ij}$, and $\rho(R_\st{S})$ is a scalar function to be determined [without weight: the additional weight $1/3$ is provided by the delta function $\delta (r-R_\st{S})$].

To determine $\rho (R_\st{S})$ we have to ask that changing the radius of the sphere 
does not change the number of particles $n$:
\begin{equation}
\!\!\!
\int \, \d \theta \d \phi \d r \sqrt h  \, \rho  \, \delta(r- R_\st{S})  =
\rho(R_\st{S}) \int \d \theta \d \phi  \sqrt{h(R_\st{S} )} = 4 \pi \, n  \,,
\end{equation}
which fixes $\rho(R_\st{S})$. Now we can rescale the momentum $n \, p_r =  P_\st{S}$, and the rest mass $n \, m_0 = M$ of the single particle
into the momentum and the mass of the whole shell, so that $n$ drops out of the equations.
Now we can integrate over $d\theta d\phi$, 
\begin{equation}
\begin{aligned}
\!\!\!\!\!\!\!\!& \int \mathcal H \, d\theta d\phi   = - 4 \pi \, \delta(r-R_\st{S}) \sqrt{g^{rr} \, P_\st{S}^2 +M^2 }  \,, \\
\!\!\!\!\!\!\!\!& \int  \mathcal H_i \,  d\theta d \phi  =  4 \pi \, \delta^r{}_i \delta (r-R_\st{S}) \, P_\st{S} \,,
\end{aligned}
\end{equation}
and the three constraints~(\ref{SphericalADMconsts}), are modified by the addition of the thin shell into
\begin{equation}\label{ConstraintsThinShell}\!\!\!
\begin{aligned}
&\textstyle -\frac{1}{6 \sigma \mu ^2} \left[ \sigma ^2 \mu s^2 +  4 f^2 \mu ^3 -4 f \sigma \mu ^2 s + 12 \sigma \mu \sigma ''   - 12 \sigma \sigma ' \mu ' 
 \right. 
\\
&\textstyle \left. ~~~~- 3 \mu (\sigma')^2  - 12 \sigma \mu ^3   \right]
= \textstyle\delta(r-R_\st{S})  \sqrt{\frac{P_\st{S}^2}{\mu^2} +M^2}   \,,
\\
&\textstyle \mu f' - {\sfrac 1 2} s \sigma' = - {\frac {P_\st{S}} 2}  \delta(r-R_\st{S})  \,, 
\qquad \textstyle \mu f + s \sigma = 0 \,.
\end{aligned}
\end{equation}

\subsection{Jump conditions}

After solving wrt $s$ the maximal-slicing constraint, $ s  = - \frac{\mu}{\sigma} f$, we can rewrite the  second constraint in~(\ref{ConstraintsThinShell}) as
\begin{equation} \label{DiffeoConstraintWithShell}
\frac{\mu}{\sqrt \sigma} \left(  f \sqrt \sigma  \right)'  = - {\sfrac 1 2}  P_\st{S}    \delta(r-R_\st{S}) \,.
\end{equation}
The above equation has the form
\begin{equation}
F'(r) = G(r) \, \delta (r - r_0) \equiv G(r_0)  \, \delta (r - r_0) \,,
\end{equation}
in any open set which does not include $r_0$ the solution to such an equation is $F(r) = \text{\it const}$. But because of the delta function on the right-hand side we cannot assume the continuity of $F$. In fact, integrating the equation from $r=0$ to $r$ one gets
\begin{equation}
F(r) - F(0) = G(r_0) \, \Theta (r - r_0) + \text{\it const.} \,,
\end{equation}
where $ \Theta (x) = \left\{ \begin{aligned}
&0 \,, ~ x<0
\\
&1 \,, ~ x>0
\end{aligned}\right.$ is the Heaviside distribution. Alternatively we can write
\begin{equation}\label{FormaEqDiffconDelta}
\begin{aligned}
&F(r) = F_- \, \Theta (r_0-r) +F_+ \, \Theta (r - r_0) \,,
\\
&F_+ - F_- = G(r_0) \,.
\end{aligned}
\end{equation}
 So~(\ref{DiffeoConstraintWithShell}) is solved by
\begin{equation} \label{SolutionDiffeoForf}
  f (r) =  \frac{A_-}{\sqrt{\sigma}}  \Theta(R_\st{S}-r)+ \frac{A_+}{\sqrt{\sigma}}  \Theta(r-R_\st{S}) \,,
\end{equation}
with the `jump condition'
\begin{equation}\label{DeltaAjumpcondition}
 A_+ - A_- =  - \frac{\sigma^{\frac 1 2}(R_\st{S}) }{2\mu(R_\st{S})} P_\st{S}  \,.\end{equation}
For the Hamiltonian constraint, we should look at Eq.~(\ref{ConstraintsThinShell}) 
and check which terms on the left-hand side can be divergent at $r = R_\st{S}$. $f$ and $s$ are not derived, and therefore they can contribute at most with a theta function, but not give any Dirac delta. $\mu$ and $\sigma$ have to be continuous because they are components of the metric.  Therefore their first derivatives $\mu'$ and $\sigma'$ can be at most discontinuous but not divergent, like $f$ and $s$. The second derivatives $\mu''$ and $\sigma''$, however, can be divergent, if the first derivatives are discontinuous. The only second derivative that appears is that of $\sigma$, so we can write
\begin{equation}\label{SingularPartOfSigma''} \textstyle
\text{\it singular part of}\left( \frac{2 \, \sigma''}{\mu } \right) =  - \delta(r-R_\st{S})   \sqrt{\frac{P_\st{S}^2}{\mu^2} +M_\st{S}^2}  \,.
\end{equation}
So we have to assume that
$\sigma$ is continuous, its first derivative has a jump, and its second derivative produces a
Dirac delta term. The prototype of one such  function is
\begin{equation}
y(r) = y_1(r) + (y_2(r) - y_2(R_\st{S})) \Theta (r-R_\st{S}) \,,
\end{equation}
where $y_1(r)$ and $y_2(r)$ are continuous functions. Taking its second derivative:
\begin{equation}
\begin{aligned}
y'' =& y_1'' + y_2''  \Theta (r-R_\st{S})+2 y_2'  \delta (r-R_\st{S})
\\
&+ (y_2 - y_2(R_\st{S})) \delta' (r-R_\st{S}) \,,
\end{aligned}
\end{equation}
a distribution of the form $z(r) \delta' (r-R_\st{S}) $ is not simply  $z(R_\st{S}) \delta' (r-R_\st{S}) $:  
with a smearing it is easy to show that it is equivalent to the distribution $-z'(r) \delta (r-R_\st{S}) + z(R_\st{S}) \delta' (r-R_\st{S})$, and in our case $z(R_\st{S}) = \lim_{r\to R_\st{S}}   (y_2(r) - y_2(R_\st{S})) = 0$. Then the above equation reads
\begin{equation}
y'' = y_1'' + y_2''(r)  \Theta (r-R_\st{S})+ y_2'(R_\st{S})  \delta (r-R_\st{S}) \,.
\end{equation}
In terms of $y$, it is easy to see that the divergent term in $y''(r)$ 
can be written as $y_2'(R_\st{S})  \delta (r-R_\st{S}) =  \left( \lim_{r\to R_\st{S}^+}  y'(r)  - \lim_{r\to R_\st{S}^-}  y'(r)  \right)  \delta (r-R_\st{S}) $. Then the jump condition for $\sigma''$ can be written as
\begin{equation} \label{lambdaJump}
\!\!\!\!\begin{aligned}
\sigma '' =& \left( \lim_{r\to R_\st{S}^+}  \sigma'(r)  - \lim_{r\to R_\st{S}^-}  \sigma'(r)  \right)   \delta  (r-R_\st{S}) 
+ \text{\it regular part.}
\end{aligned}
\end{equation}
This (times $2/\mu$) is the only divergent part of Eq.~(\ref{SingularPartOfSigma''}), and therefore we can identify it with the right-hand side:
\begin{equation}
 \lim_{r\to R_\st{S}^+}  \sigma'  - \lim_{r\to R_\st{S}^-}  \sigma'  = - {\sfrac 1 2} \sqrt{P_\st{S}^2 + M_\st{S}^2 \mu^2(R_\st{S})} \,,
\end{equation}
to produce our second jump condition.

It is convenient, at this point, to define some quantities which will appear in all the jump conditions below:
\begin{equation} \label{EqgttDotAtR}
 \lim_{r\to R_\st{S}^+}  \sigma' = \gamma \,, ~~   \lim_{r\to R_\st{S}^-}  \sigma' = \kappa \,, ~~ \sigma(R_\st{S}) = \rho^2 \,,
\end{equation}
whatever diffeo gauge we choose, around $r=R_\st{S}$,  $\sigma(r)$ can be written as
\begin{equation}
\begin{aligned}
\sigma =& \rho^2 +  (r-R_\st{S})  \gamma \, \Theta(r-R_\st{S}) \\
&+(r-R_\st{S})  \kappa \, \Theta(R_\st{S}-r) 
+ \mathcal O \left[ (r-R_\st{S})^2\right] \,,
\end{aligned}
\end{equation}
then, from the expression above, it is easy also to deduce that
\begin{equation}
 \lim_{r\to R_\st{S}^+} \dot  \sigma  = 2 \rho \dot \rho -  \gamma \dot R_\st{S}  \,, ~~  \lim_{r\to R_\st{S}^-} \dot  \sigma = 2 \rho \dot \rho -  \kappa \dot R_\st{S}  \,.
\end{equation}

We are now in position to demand the continuity of $\mu$. Its expressions inside and 
outside of the shell do not coincide:
\begin{equation}
\mu = \left\{ \begin{array}{ll} 
 |\sigma'|/ \sqrt{\frac{A_-^2}{\sigma } - 8 \, \mass_-  \, \sqrt{\sigma}  + 4 \, \sigma } \,, &  r<R_\st{S}
\\
 |\sigma'| / \sqrt{\frac{A_+^2}{\sigma }  - 8 \, \mass_+  \, \sqrt{\sigma}  + 4 \, \sigma} \,, & r>R_\st{S}
\end{array} \right.
\end{equation}
and we have to demand that the left and right limit of $\mu$ coincide:
\begin{equation}
 \lim_{r\to R_\st{S}^+} \mu(r) =  \lim_{r\to R_\st{S}^-} \mu(r) \,,
\end{equation}
that is, 
\begin{equation}\label{ContinuityOfMu}
\!\!\!  \frac{|\kappa|}{\sqrt{ \left(\frac {A_-} {2 \rho^{2}}\right)^2  - \frac{2 \mass_-}{\rho} +  1  } }= \frac{|\gamma|}{\sqrt{ \left(\frac {A_+  } {2 \rho^{2}}\right)^2  - \frac{2 \mass_+}{\rho} +  1      } } \,.
\end{equation}
This is a new equation we have to take into account, together with the jump conditions above,
which, in the new notation, can be written as 
\begin{equation}\label{JumpGammaKappa}
\gamma  - \kappa   = - {\sfrac 1 2} \sqrt{P_\st{S}^2 + M_\st{S}^2  \mu^2(R_\st{S}) } \,.
\end{equation}

We can completely eliminate $\gamma$ and $\kappa$ from the two jump conditions~(\ref{ContinuityOfMu}) and~(\ref{JumpGammaKappa}): using Eq.~(\ref{DeltaAjumpcondition}) into Eq.~(\ref{JumpGammaKappa}) 
\begin{equation}\label{JumpGammaKappa2} 
\frac{\kappa}{| \mu(R_\st{S}) |}  - \frac{\gamma}{| \mu(R_\st{S}) |}  =  \sqrt{\frac{(A_+ - A_-)^2}{\rho_\st{S}^2} + {\sfrac 1 4} M_\st{S}^2 } \,,
\end{equation}
by taking twice the square of the above equation, we can make it independent of the signs of $\kappa$ and $\gamma$,
\begin{equation}\label{JumpGammaKappa3} 
\left(\frac{\kappa^2}{\mu^2(R_\st{S})}  + \frac{\gamma^2}{\mu^2(R_\st{S})} -   \frac{(A_+ - A_-)^2}{\rho_\st{S}^2} -  {\sfrac 1 4} M_\st{S}^2 \right)^2 = \frac{4\,\kappa^2\gamma^2}{\mu^4(R_\st{S})} \,.
\end{equation}
Now, using the definition of $\mu(r)$ at $r= R_\st{S}$:
\begin{equation}
\left\{ \begin{aligned}
&\textstyle 
\frac{\gamma^2}{\mu^2(R_\st{S})}  = \left(\frac {A_+}{\rho}\right)^2  -8 \, \mass_+ \, \rho + 4  \rho^2 
\\
&\textstyle 
\frac{\kappa^2}{\mu^2(R_\st{S})} = \left(\frac {A_- } {\rho}\right)^2  - 8 \, \mass_- \, \rho +4 \rho^2 
\end{aligned}\right. \,,
\end{equation}
 we end up with the following `on-shell condition':
\begin{equation}\label{OnshellCondition}
\begin{aligned}
&\textstyle
\left( \frac {A_+ \, A_-}{\rho^2} - 4 (\mass_+ + \mass_-) \, \rho + 4 \rho^2  -  {\sfrac 1 8} M_\st{S}^2 \right)^2 =
\\
&\textstyle \left(\frac {A_+^2}{\rho^2} - 8 \, \mass_+ \, \rho + 4  \rho^2    \right) \left(  \frac{A_-^2} {\rho^2}  - 8 \, \mass_- \, \rho +4 \rho^2   \right) \,.\end{aligned}
\end{equation}

\subsection{Symplectic structure}

In order to discuss the dynamics of the system, we need to know which of the reduced-phase-space variables are canonically conjugate to each other. In other words, we need to calculate the symplectic form. By definition, the conjugate variables of the extended phase space are $g_{ij}$ and $p^{ij}$, as well as $R_\st{S}$ and $P_\st{S}$. Therefore the pre-symplectic potential is
\begin{equation}
\theta = \int \d r \, \d \theta \, \d \phi \, p^{ij} \, \delta g_{ij} + 4 \pi  \, P_\st{S} \delta R_\st{S} \,,
\end{equation}
restricting it through spherical symmetry and integrating in $\d\theta \d\phi$:
\begin{equation}
\theta = 4 \pi  \int_0^\infty \d r  \left( 2 f \, \delta \mu + s \, \delta \sigma\right)+ 4 \pi \, P_\st{S} \delta R_\st{S} \,.
\end{equation}
Now we may impose the maximal-slicing constraint~$\mu \, f =  - s \, \sigma$, and the solution to the diffeo constraint~(\ref{SolutionDiffeoForf}),
\begin{equation}
\begin{aligned}
&\theta = 4 \pi  \int_0^\infty  \d r  \left( 2 f \, \delta \mu -\frac{ \mu \, f}{\sigma} \delta \sigma \right) + 4 \pi \, P_\st{S} \delta R_\st{S}  
\\
&= - 4 \pi  \int_0^\infty \d r \frac{2 \mu}{\sqrt{\sigma}} \delta (f \sqrt \sigma)+ 4 \pi \, P_\st{S} \delta R_\st{S}  
\,.
\end{aligned}
\end{equation}
now, using Eq.~(\ref{DeltaAjumpcondition}) we observe that the first term in the last equation cancels the second term:
\begin{equation}
\begin{aligned}
 &- 4 \pi  \int_0^\infty \d r \frac{2 \mu}{\sqrt{\sigma}} \delta (f \sqrt \sigma) =
\\
& - 8 \pi  \int_0^\infty \frac{\mu}{\sqrt \sigma} \delta \left[  A_-  \, \Theta(R_\st{S}-r) +  A_+  \, \Theta(r-R_\st{S})  \right]
\\
=&
 -  8 \pi  \left[ \delta  A_-  \int_0^{R_\st{S}} \d r  \frac{\mu}{\sqrt \sigma}  + \delta  A_+  \int_{R_\st{S}}^{\infty} \d r  \frac{\mu}{\sqrt \sigma}   \right]
\\ 
&+ 8 \pi \left[ (A_+  - A_-)  \frac{\mu(R_\st{S})}{\sqrt{\sigma(R_\st{S})}} \delta R_\st{S}\right] 
\\
=& -  8 \pi  \left[ \delta  A_-  \int_0^{R_\st{S}} \d r  \frac{\mu}{\sqrt \sigma}  + \delta  A_+   \int_{R_\st{S}}^{\infty} \d r  \frac{\mu}{\sqrt \sigma}  \right]
\\
&
- 4 \pi \, P_\st{S} \, \delta R_\st{S}   \,.
\end{aligned}
\end{equation}
And therefore the symplectic potential reduces to
\begin{equation}\label{SingleShellSymplectic1}
\theta =  
-  8 \pi  \left[  \delta  A_-  \int_0^{R_\st{S}}   \frac{\mu \d r}{\sqrt \sigma}  +   \delta  A_+  \int_{R_\st{S}}^{\infty}   \frac{\mu \d r}{\sqrt \sigma} \right]  \,.
\end{equation}

%

\section{Thin shell in an asymptotically flat nonexpanding region}\label{sec:fullsystem}

So far we have kept everything as general as possible. We now need to specialize to a particular model by fixing some of the integration constants.

\subsection{Boundary conditions}

The manifold we are studying has two boundaries. one at $r \to \infty$ (asymptotic infinity) and one at the origin $r \to 0$. At infinity, as we discussed in the introduction, we have to set $A_+=0$, in order to be consistent with~\cite{Birkhoff_SD}. This might not be entirely physically justified from the perspective of SD~\cite{BirkhoffFlavio}, but but here we are focused on determining whether the gravitational collapse of our thin shell of dust can generate the solution of~\cite{Birkhoff_SD}, and therefore we have to impose the same conditions at infinity. Moreover, it should be noted that these are the  standard  asymptotically-flat conditions in GR \cite{ReggeTeitelboim, beig1987poincare}.

Regarding the inside of the shell, we cannot assume $A_- =0$ because that would trivialize the dynamics (if $A_+ = A_- =0$ then $P_\st{S}=0$). But we cannot assume $A_-\neq 0$ for the entire interior either, because the metric would then develop a singularity, or a `piercing'-like defect (see~\cite{BirkhoffFlavio}). We have then to assume that there is some other matter inside the shell, whose expansion compensates $A_-$ and puts the effective value of the integration constant $A$ at the origin to zero. 
A realistic model of such matter could be, for example, a homogeneous-density star which accretes our thin shell as an additional layer.
The price to pay is that $\mass_-$ inside the shell cannot be put to zero either, otherwise we would be assuming the existence of some kind of matter with nonzero momentum but vanishing mass-energy.  We thus set  $\mass_-$ as a free parameter of our model.
We will ignore  the dynamics of this conjectured matter near the origin, and concentrate on the exterior of the shell. The integration constants $\mass_+$ and $\mass_-$ are conserved quantities which characterize our solutions as freely adjustable parameters of the model. The only thing we assume about them is their positivity, because they are related to the Misner--Sharp mass and its positivity follows from the dominant energy condition when $A_+ = 0$~\cite{BirkhoffFlavio}.\footnote{Although the condition is borrowed from the space-time picture, it should be valid in the appropriate limits.}

%
%

\subsection{Phase space}

Eq.~(\ref{SingleShellSymplectic1}), using the isotropic gauge $\mu = \sqrt \sigma/ r$, becomes
\begin{equation}
\theta =   8 \pi  \,  \log R_\st{S} \, \delta \left( A_+ - A_- \right)   + \text{\it exact form}  \,,
\end{equation}
and, recalling Eq.~(\ref{DeltaAjumpcondition}), $ A_+ - A_- =  -  {\frac 1 2}  P_\st{S} \, R_\st{S}   $, we get (modulo an exact form)  $\theta =   - 4 \pi  P_\st{S}  \,  \log  R_\st{S} \,    \delta  R_\st{S}  - 4 \pi  R_\st{S}  \log  R_\st{S} \,   \delta   P_\st{S}$, which gives the following canonical symplectic form:
\begin{equation}\label{SingleShellSymplectic2}
\omega = \delta \theta =    4 \pi  ~ \delta  P_\st{S} \wedge   \delta   R_\st{S}    \,,
\end{equation}
so, in this gauge, $ P_\st{S} $ and  $ R_\st{S}  $ are canonically conjugate. 

%
%

\subsection{Exact solution of the constraints}

So we set $A_+=0$.  Then in the region outside the shell we can use `isotropic' coordinates, $\mu^2 = \frac{\sigma}{r^2}$, so that the metric outside is conformal to the Euclidean metric:
\begin{equation}
\d s^2 = \mu^2 \left[ \d r^2  + r^2\left( \d \theta^2 +  \sin^2 \theta \d \phi^2 \right) \right] \,.
\end{equation}
Using this gauge, Eq.~(\ref{VacuumSolutionForMu}) can be treated as a differential equation for $\sigma$:
\begin{equation}\label{SingleShell_IsotropicEquation}
\frac{(\sigma')^2}{\frac{A_+^2}{\sigma} - 8 \, \mass_+ \sqrt{\sigma} +4 \sigma} = \frac{\sigma}{r^2} \,,
\end{equation}
and, using $A_+=0$, we can integrate this equation as
\begin{equation}
 \left(\frac{2 \sqrt{\sigma} - 2\,\mass_+ + 2 \sqrt{\sigma -2\,\mass_+ \sqrt{\sigma}}}{2\,\mass_+ \, k_+} \right)^2  =   \left( \frac r {2 \,\mass_+} \right)^{\pm 1}  \,,
\end{equation}
where $k_+$ is a positive integration constant. Solving for $\sigma$:
\begin{equation}
\sigma  = \frac{\mass_+^2}{4}  \left[ \sqrt{k_+} \, \left( \frac r {2 \,\mass_+} \right)^{\pm {\frac 1 2} }  + \frac 1 {\sqrt k_+} \left( \frac {2 \,\mass_+ } r \right)^{\pm {\frac 1 2} }  \right]^4\,.
\end{equation}
an explicit calculation immediately shows that the above expression is identical whichever sign we choose (modulo a transformation $k_+\to 1/k_+$), so we can write
\begin{equation}\label{SingleShellSigma}
\sigma  = \frac{\mass_+^2}{4}  \left[ \left( \frac {k_+\, r} {2 \, \mass_+} \right)^{\frac 1 2}  +  \left( \frac {2 \, \mass_+ }{k_+ \, r} \right)^{\frac 1 2}  \right]^4\,.
\end{equation}
The minimum of $\sigma$ is always $4 \, \mass_+^2$, which is where $\sqrt{\sigma}/\mass_+$ reaches the only zero of the polynomial $\mathscr P(\sqrt\sigma /\mass_+)$. This minimum is at the coordinate radius $r = 2 \mass_+ /k_+$. The integration constant $k_+$ has the only role of rescaling the coordinate $r$ by a constant factor, and it is therefore an effect of a residual radial diffeomorphism redundancy. We can fix this redundancy by imposing $\sigma \xrightarrow[r \to \infty]{} r^2$, which means $k_+ =4$.

%

\subsection{Solution of the jump conditions}

\begin{figure}[t!]\label{OnShellPlots}
\center
\includegraphics[width=0.35\textwidth]{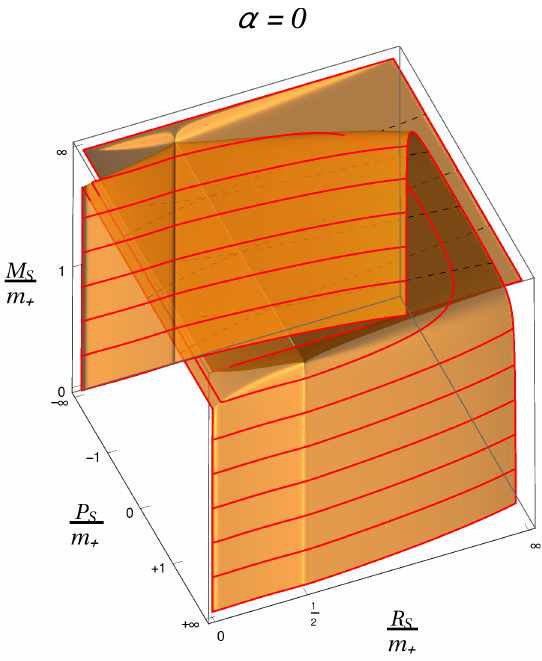}
~~~~~~~~
\includegraphics[width=0.35\textwidth]{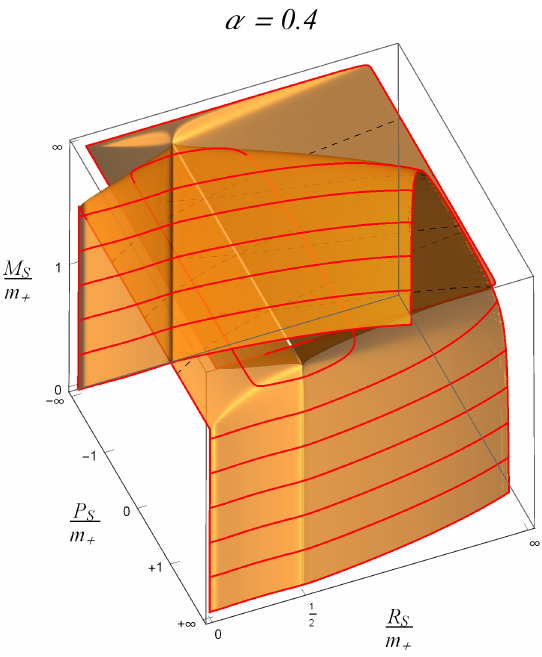}
\\
\includegraphics[width=0.35\textwidth]{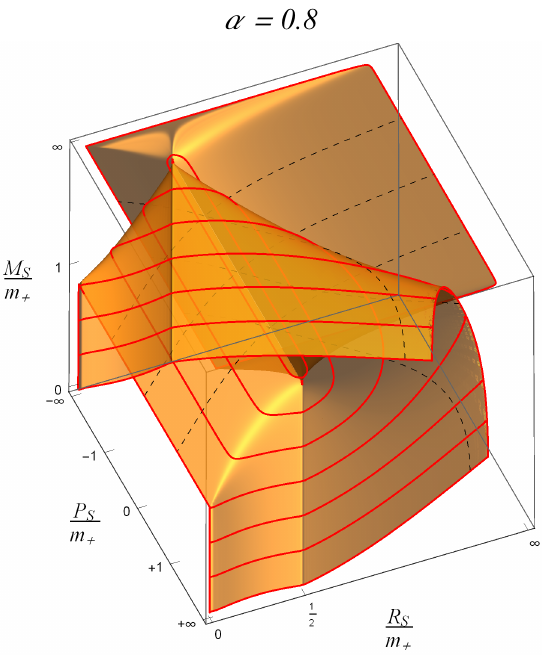}
~~~~~~~~
\includegraphics[width=0.35\textwidth]{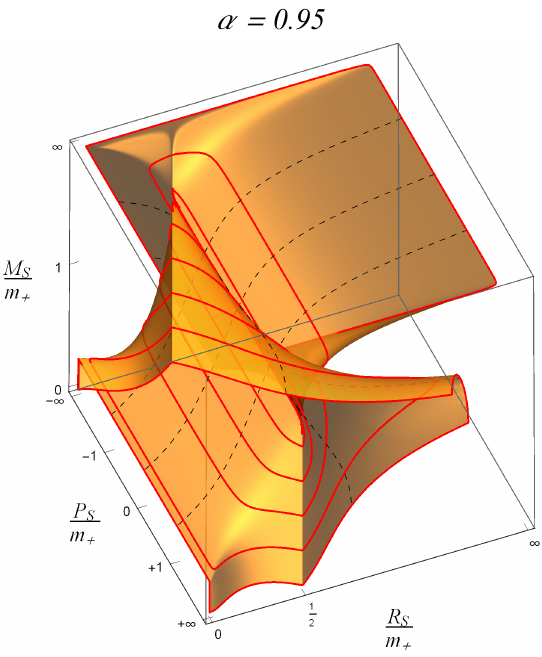}
\caption{\it
Plots of the on-shell surface~(\ref{OnShell_dimensionless}) in the space $R \in [0,\infty)$, $P \in (-\infty,\infty)$,  $M \in [0,\infty)$, and for certain fixed values of $\alpha = m_- / m_+ \in  [0,1)$.  The three variables $R$, $P$ and $M$ have been compactified by taking their $\arctan$. The red curves represent the constant-$M$ cross-sections, which are on-shell curves in the phase space $R$, $P$. Notice how all the curves `bounce' on the $P = \pm \infty$ boundary of phase space at $R = 1/2$. }
\end{figure}

The solution~(\ref{SingleShellSigma}) is valid outside of the shell.  Inside the shell $\sigma$ will be different, because $A_- \neq 0$.  However in this region we cannot analytically solve Eq.~(\ref{VacuumSolutionForMu}) in isotropic gauge. Whatever the solution turns out to be, it will depend on one integration constant $k_-$.  We need to satisfy two conditions, Eq.~(\ref{ContinuityOfMu}), imposing the continuity of $\mu$ (which, because we are working in isotropic gauge, implies also the continuity of $\sigma$), and Eq.~(\ref{JumpGammaKappa}). The two equations depend on the left- and right- derivatives of $\sigma$ at the shell, $\kappa$ and $\gamma$, which are in turn determined by the two integration constants $k_-$ and $k_+$. 
We can assume that $k_-$ has been solved by one of the two equations ~(\ref{ContinuityOfMu}) and~(\ref{JumpGammaKappa}), and the other independent condition will be Eq.~(\ref{OnshellCondition}), which, after imposing $A_+ = 0$ and $ A_- =   \frac 1 2 R_\st{S} \,P_\st{S} $ (valid in isotropic gauge), is:
\begin{equation}\label{OnshellCondition2}
\begin{aligned}
&\textstyle
\left( - 4 (\mass_+ + \mass_-) \, \rho + 4 \, \rho^2  -  {\sfrac 1 8} M_\st{S}^2 \right)^2 =
\\
&\textstyle \left(4  \rho^2 - 8 \, \mass_+ \, \rho   \right) \left(  \frac{R_\st{S}^2 \,P_\st{S}^2} {4 \rho^2}  - 8 \, \mass_- \, \rho +4 \rho^2   \right) \,.\end{aligned}
\end{equation}
The equation above depends on $R_\st{S}$ through $\rho = \sqrt{\sigma(R_\st{S})}$. 
To further reduce the number of parameters, we express everything in units of $m_+$:
\begin{equation}
R_\st{S} = \mass_+ \, R \,, ~~ P_\st{S} = \mass_+ \, P  \,, 
~~
m_- = \mass_+  \, \alpha \,, ~~ M_\st{S} =  \mass_+ \, M \,,
\end{equation}
then Eq.~(\ref{OnshellCondition2}) becomes
\begin{equation}\label{OnShell_dimensionless} 
\textstyle  \frac{M^4}{64}  + \frac{(2 R+1)^4}{R^2} \left( (\alpha -1)^2 - M^2 \frac{\left(4 R^2-4 \alpha  R+1\right)}{16 (2 R+1)^2}\right) =   \frac{(1-2 R)^2 R^2}{(2 R+1)^2}  P^2\,,
\end{equation}
In Fig.~\ref{OnShellPlots} we plot the on-shell curves $P$ vs. $R$, for any possible choice of rest-mass $M$, and for a set of choices of $\alpha$. Notice that the constant $\alpha$, on physical grounds, should be smaller than one (and larger than zero), as the ADM mass inside the shell should be smaller than outside.

We conclude this section with an analysis of the Hamiltonian vector flow in reduced phase-space $(R_\st{S},P_\st{S})$.  Consider Eq.~(\ref{OnshellCondition2}) as a condition on $m_+$: if we take into account the definition of $\rho = \sqrt{\sigma(R_\st{S})}$, it turns into an eight-order equation for $m_+$. Let us write it as $F(m_+,R_\st{S},P_\st{S};M,m_-) = 0$. Its solution gives the ADM energy  $m_+$ as a function of the dynamical variables $R_\st{S}$ and $P_\st{S}$ (as well as the constant parameters $m_-$ and $M$). This is the Hamiltonian generator of evolution in maximal-slicing time. If all we are interested in are the equations of motion of $R_\st{S}$ and $P_\st{S}$ in this time variable, we can avoid having to explicitly solve $F=0$.  We can instead differentiate $F$ wrt  all of the dynamical variables: $\frac{\partial F}{\partial m_+} d m_+ + \frac{\partial F}{\partial R_\st{S}} d  R_\st{S} + \frac{\partial F}{\partial  P_\st{S}} d P_\st{S} = 0 $, which implies that $\frac{\partial m_+ }{\partial R_\st{S}} = -\left.  \frac{\partial F}{\partial R_\st{S}} / \frac{\partial F}{\partial m_+} \right|_{F =0} $ and $\frac{\partial m_+ }{\partial P_\st{S}} = - \left.  \frac{\partial F}{\partial P_\st{S}}/ \frac{\partial F}{\partial P_+} \right|_{F =0}$.
Then the Hamiltonian equations of motion generated by $m_+$ are
\begin{equation}\label{HamiltonianEquationsOfMotion} \textstyle
\dot R_\st{S} =  - \left. \frac{\partial F}{\partial P_\st{S}} \left(\frac{\partial F}{\partial P_+} \right)^{-1}\right|_{F =0} \,,
\qquad
\dot P_\st{S} = \left. \frac{\partial F}{\partial R_\st{S}} \left( \frac{\partial F}{\partial m_+} \right)^{-1} \right|_{F =0}\,,
\end{equation}
which, before replacing the solution $F=0$, are two perfectly tractable functions of $R_\st{S}$, $P_\st{S}$ and $m_+$.

Eq.~(\ref{HamiltonianEquationsOfMotion}) allows us to  study the Hamiltonian vector flow in phase space. In particular we can check its behaviour at the `throat' $R_\st{S} \to m_+/2$ without having to solve $F=0$. It turns out that the vector flow vanishes at the throat:
\begin{equation}
\dot R_\st{S}  \xrightarrow[R_\st{S} \to m_+/2]{} 0 \,,
\qquad
\dot P_\st{S}   \xrightarrow[R_\st{S} \to m_+/2]{} 0\,.
\end{equation}
The limits before are the same irrespective of the direction they are taken from.
so, as expected, in maximal-slicing time the shell `freezes' at the throat. One can also prove that it takes an infinite amount of maximal-slicing time for the shell to reach the throat, by explicitly integrating the vector flow. Maximal-slicing time has no intrinsic physical meaning: one of the fundamental relational underpinnings of SD is that time should be abstracted from the change of physical (\emph{i.e.} shape) degrees of freedom. In this sense maximal slicing time is associated to the change in the DOFs of a clock far away from the origin.

\section{Outlook and Conclusions}\label{sec:conclusions}

We are now in position to give at least partial answers to the questions we set forth at the beginning. First,  does the `wormhole'-like line element found in~\cite{Birkhoff_SD} emerge from the gravitational collapse of spherically symmetric matter? Under the same assumptions of asymptotically flat boundary conditions (\emph{i.e.} $p^{ij} \to r^{-2} ~\Rightarrow ~ A_+ =0$) at infinity the answer is clearly positive. The line element given by the areal radius~(\ref{SingleShellSigma}) outside of the shell when $k_+ = 4$ is identical to that of ref.~\cite{Birkhoff_SD}, so, as it collapses, the shell leaves in its wake the wormhole line element.

\looseness=-1
The `on-shell' relation~(\ref{OnShell_dimensionless}) produces, for any value of $M= \frac{M_\st{S}}{m_+}$ and  $\alpha = \frac{m_-}{m_+}$ a curve in the $P$-$R$ space, which reaches the boundary of phase space $P 
\to \pm \infty$ at $R = \frac 1 2$, that is, $R_\st{S} = \frac {\mass_+} 2$. This value of $R_\st{S}$ coincides with the throat of the wormhole line element with mass $\mass_+$. This result implies that the collapsing shell does not reach the throat in a finite maximal-slicing time. This time parameter coincides with the experienced reading of a clock of an inertial observer at infinity. The preliminary conclusion is that the shell `freezes' at the throat and cannot be observed to cross it. However, as we know, maximal-slicing time can at best be an infinitely-thin layer of York time (the time parameter of CMC slicings). Whether the shell crosses the throat or not thus has to be postponed for the study of gravitational collapse in a cosmological setting (in which we take into account a cosmological constant, a nonzero York time and a compact spatial manifold). 

 If this behavior does arise as a limit of the cosmological setting, at this point we would offer a tentative interpretation: the ratios of scales in a closed space-time (total volume, cosmological constant, and  MS mass) may only allow for a given (non-zero) minimum areal radius (as it only allows for a maximum one). Thus either the system undergoes a bounce, or the shape degrees of freedom around the throat asymptotically  (in time) freeze with respect to other local shape degrees of freedom.  In either case, the dynamical behavior seems to be non-singular. 

For the moment, we can study the on-shell curves of~(\ref{OnShell_dimensionless}), and observe that they continue past the point $R = \frac 1 2$ where they reach the boundary of phase space.  The solution curves fall  into two topologically-distinct kinds: the closed and the open ones. The former are closed loops which touch the boundary of phase space at two points. They correspond to the cases in which the shell does not have enough kinetic energy to reach infinity, and recollapses back. Interestingly, this behaviour is observed on both sides of the throat  $R = \frac 1 2$, so the shell recollapses also when it is in the region beyond the throat.
The other kind of curves are the open ones, which reach the asymptotic boundary $R \to \infty$, and the other asymptotic infinity at $R \to 0$.

\section*{Acknowledgments}

We would like to thank Vasudev Shyam, Neill \'O Murchadha, Lee Smolin, Sean Gryb, and Daniel Guariento for discussions. FM and TK were supported by the Foundational Questions Institute (FQXi) through grant FQXi-RFP3-1339. This research was supported  by Perimeter Institute for Theoretical Physics. Research at Perimeter Institute is supported by the Government of Canada through Industry Canada and by the Province of Ontario through the Ministry of Research and Innovation.

\providecommand{\href}[2]{#2}\begingroup\raggedright\endgroup


\begin{thebibliography}{10}

\bibitem{ADM}
R.~L. Arnowitt, S.~Deser, and C.~W. Misner, ``The dynamics of general
  relativity,'' \href{http://arxiv.org/abs/gr-qc/0405109}{{\ttfamily
  arXiv:gr-qc/0405109}}. In Gravitation: an introduction to current research,
  Louis Witten ed., chapter 7, pp 227--265.

\bibitem{Birkhoff_SD}
H.~Gomes, ``A Birkhoff theorem for Shape Dynamics,'' {\em Class. Quantum Grav.}
  {\bfseries 31} (2014) 085008,
  \href{http://arxiv.org/abs/1305.0310}{{\ttfamily arXiv:1305.0310 [gr-qc]}}.

\bibitem{BirkhoffFlavio}
F.~Mercati, ``On the fate of Birkhoff's theorem in Shape Dynamics,''
  \href{http://dx.doi.org/10.1007/s10714-016-2134-2}{{\em accepted for
  publication by Gen. Rel. Grav.} (2016) },
  \href{http://arxiv.org/abs/1603.08459}{{\ttfamily arXiv:1603.08459 [gr-qc]}}.

\bibitem{ThroughTheBigBang}
T.~A. Koslowski, F.~Mercati, and D.~Sloan, ``Relationalism Evolves the Universe
  Through the Big Bang,'' \href{http://arxiv.org/abs/1607.02460}{{\ttfamily
  arXiv:1607.02460 [gr-qc]}}.

\bibitem{FlaviosSDtutorial}
F.~Mercati, {\em Shape Dynamics: Relativity and Relationalism}.
\newblock Oxford Univ. Press, 2016.
\newblock \href{http://arxiv.org/abs/1409.0105}{{\ttfamily arXiv:1409.0105
  [gr-qc]}}.
\newblock (preliminary version on the arXiv).

\bibitem{Tim_Proceedings_TheoryCanada9}
T.~Koslowski, ``The shape dynamics description of gravity,''
  \href{http://dx.doi.org/10.1139/cjp-2015-0029}{{\em Can. J. Phys.} {\bfseries
  93} (2015) 956--962}, \href{http://arxiv.org/abs/1501.03007}{{\ttfamily
  arXiv:1501.03007 [gr-qc]}}.

\bibitem{gryb:shape_dyn}
H.~Gomes, S.~Gryb, and T.~Koslowski, ``Einstein gravity as a 3D conformally
  invariant theory,''
  \href{http://dx.doi.org/10.1088/0264-9381/28/4/045005}{{\em Class. Quant.
  Grav.} {\bfseries 28} (2011) 045005},
  \href{http://arxiv.org/abs/1010.2481}{{\ttfamily arXiv:1010.2481 [gr-qc]}}.

\bibitem{Gomes:linking_paper}
H.~Gomes and T.~Koslowski, ``The Link between General Relativity and Shape
  Dynamics,'' \href{http://dx.doi.org/10.1088/0264-9381/29/7/075009}{{\em
  Class. Quant. Grav.} {\bfseries 29} (2012) 075009},
  \href{http://arxiv.org/abs/1101.5974}{{\ttfamily arXiv:1101.5974 [gr-qc]}}.

\bibitem{MisnerSharpMass}
C.~W. Misner and D.~H. Sharp, ``Relativistic Equations for Adiabatic,
  Spherically Symmetric Gravitational Collapse,''
  \href{http://dx.doi.org/10.1103/PhysRev.136.B571}{{\em Phys. Rev.} {\bfseries
  136} (1964) B571--B576}.

\bibitem{ReggeTeitelboim}
T.~Regge and C.~Teitelboim, ``Role of surface integrals in the Hamiltonian
  formulation of general relativity,'' {\em Annals of Physics} {\bfseries 88}
  no.~1, (1974) 286--318.

\bibitem{beig1987poincare}
R.~Beig and N.~{\'O}~Murchadha, ``The Poincar{\'e} group as the symmetry group
  of canonical general relativity,'' {\em Annals of Physics} {\bfseries 174}
  no.~2, (1987) 463--498.

\end{thebibliography}
\end{document}